\documentclass[useAMS,letter]{emulateapj}

\usepackage{amssymb}
\usepackage{natbib}
\usepackage{graphicx}

\shorttitle{Turbulence and Triggered Star Formation}
\shortauthors{Gritschneder et al.}

\begin{document}


\title{Driving Turbulence and Triggering Star Formation by Ionizing Radiation}
\author{Matthias Gritschneder$^{1}$, Thorsten Naab$^{1}$, Stefanie
  Walch$^{1}$, Andreas Burkert$^{1}$, Fabian Heitsch$^{2}$}
\affil{\\$^1$ Universit\"ats-Sternwarte M\"unchen, Scheinerstr.\ 1,
  D-81679 M\"unchen, Germany \\ $^{2}$Department of Astronomy,
  University of Michigan, Ann Arbor, MI 48109-1042, U.S.A.}
 \texttt{gritschm@usm.uni-muenchen.de}\\



\begin{abstract}

We present high resolution simulations on the impact of ionizing
radiation of massive O-stars on the surrounding turbulent interstellar
medium (ISM). The simulations are performed with the newly developed
software iVINE which combines ionization with smoothed particle
hydrodynamics (SPH) and gravitational forces. We show that radiation
from hot stars penetrates the ISM, 
efficiently heats cold low density gas
and amplifies over-densities seeded by the initial turbulence. The
formation of observed pillar-like structures in star
forming regions (e.g. in M16) can be explained by this scenario. At the
tip of the pillars gravitational collapse can be induced,
eventually leading to the formation of low mass stars.  Detailed
analysis of the evolution of the turbulence spectra shows that
UV-radiation of O-stars indeed provides an excellent mechanism to sustain and
even drive turbulence in the parental molecular cloud.
\end{abstract}




\section{Introduction}
Some of the most spectacular structures in the molecular ISM are
observed in the vicinity of hot O/B-stars or associations, e.g.  the
Horsehead nebula (B33), the three pillars of creation in M16 and the
Elephant trunk (BRC37) in IC1396. For the pillars in M16 
\citet{2002ApJ...565L..25S} find a
head to tail structure with the denser head pointing towards the OB
stars of NGC661. In addition, young stellar objects (YSOs) are present
at the tips of the pillars. In the Horsehead nebula
\citet{2006MNRAS.369.1201W} report two 
core-like structures that might undergo subsequent gravitational
collapse. Very 
recent observations by \citet{2008AJ....135.2323I} report several YSOs
close to the tip of BRC37. As a common feature these pillar-shaped
nebulae point towards a source of ionizing radiation and show signs of
present or future star formation at their tips. 

Up to now the precise physical processes leading to the formation of
these structures are not fully understood.
The morphologies suggest that feedback effects of UV-radiation and winds of 
massive stars play an important role in the formation of the pillars.
In addition, the radiation might have a strong impact on the overall
evolution of the parental cloud. 
Furthermore, molecular clouds are observed to be highly turbulent
structures. 
There is evidence that this turbulence can support the clouds
against gravitational collapse and thereby control star formation. 
As hydrodynamic and MHD turbulence decays rather quickly, the only way
to explain this high level of turbulence
would be to drive the turbulence - either on large
scales by i.e. supernova explosions or on small scales from within the
cloud by stellar outflows, winds or ionization (see
e.g. \citealt{2004ARA&A..42..211E} and \citealt{2004RvMP...76..125M}, for
reviews). The possibility of ionization driven turbulence
has  been indicated by e.g. semi-analytic models of \cite{2006ApJ...653..361K}.
In this Letter we test the hypothesis
using high resolution numerical simulations with the newly developed
code iVINE \citep[][ hereafter G08]{2009MNRAS.393...21G}.

On the theoretical side progress has been made since
\citet{1995ApJ...451..675E} first presented two-dimensional, grid-based
simulations showing that the expansion of an HII region into
the surrounding ISM can trigger star formation by sweeping up the cold
material. This is called 'collect and collapse'. Another proposed
scenario is the 'radiation driven implosion', where preexisting
density structures are driven into collapse (see
e.g. \citealt{1989ApJ...346..735B}, 
\citealt{2003MNRAS.338..545K} and G08).

For the numerical 
treatment of radiation in simulations several codes have been developed (see
\citealt{2006MNRAS.371.1057I} and references therein). Recent 
applications for the treatment of ionizing radiation in grid based
codes include e.g. \citet{2006ApJ...647..397M} and \citet{2007ApJ...671..518K}.
In SPH-codes implementations have been presented
by \citet{2005MNRAS.358..291D},
\citet{2008MNRAS.389..651P} and
\citet{2008MNRAS.386.1931A}. Simulations by
\citet{2007MNRAS.377..535D} show that ionizing radiation can slightly enhance
the formation of cores in a globally unbound molecular cloud of $10^4
\mathrm{M}_\odot$. With their choice of initial conditions the positive
feedback, the additional or faster formation of cores, outweighs the
negative feedback, the disruption of cores.
All these applications calculate the effect of a point source on the
surrounding medium, thereby focussing much more on the global effect of
the ionization. However, neither the detailed morphology of the gas
nor the impact of the ionizing radiation on the turbulence has been
investigated so far. 

\section{Initial Conditions}
\label{simulations}
We set up a box of gas with sides $4$pc long at a temperature of
$T_{cold} = 10$K and a mean number density of $\bar{n} = 300
\textrm{cm}^{-3}$, which resembles a slightly denser part of a molecular
cloud. The gas mass in the box is $474 \mathrm{M}_\odot$ which corresponds
to $25$ Jeans masses. To mimic initial turbulence we employ
a supersonic turbulent velocity field (Mach 10) with a 
steep power-law $E(k)\propto k^{-2}$, where only the
largest modes $k=1..4$ are populated initially. This setup
is allowed to freely decay under the influence of isothermal hydrodynamics 
simulated with the tree/SPH-code VINE (\citealt{2008arXiv0802.4245W},
\citealt{2008arXiv0802.4253N}). 
The individual particle time-steps in VINE are determined by using an
accuracy parameter of $\tau_\mathrm{acc}=1.0$ and a
Courant-Friedrichs-Lewy (CFL) tolerance parameter of
$\tau_\mathrm{CFL}=0.3$. 
We also use an additional time-step
criterion based on the maximum allowed change of
the smoothing length with an accuracy parameter of $\tau_\mathrm{h}=0.15$.

\begin{figure}
  \centering{
   \plotone{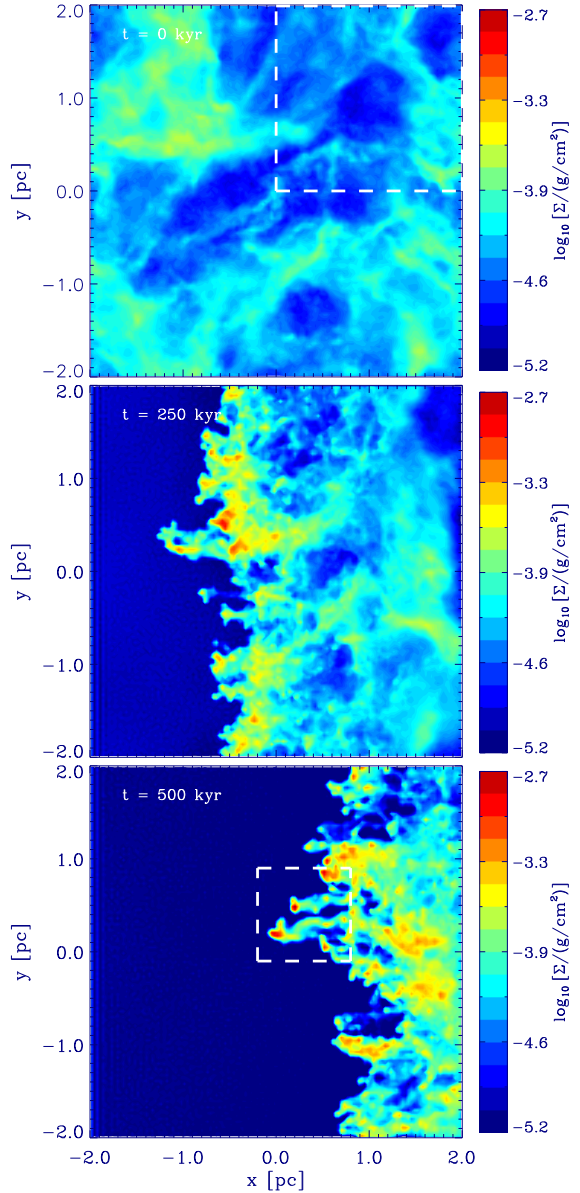}
  \caption{Evolution of the turbulent ISM under the influence of
    UV-radiation impinging form the left hand side. Color coded is the
    surface density projected along the z-direction. The time of the snapshot      is increasing from top to bottom.\label{morphology}}}
\end{figure}

After $\approx 1$Myr a Kolmogorov-like power-law
with $E(k)\propto k^{-\frac{5}{3}}$ is well established 
on all resolvable scales. The velocities now correspond to Mach 5. This
initial setup with the turbulent velocities is shown in
Fig. \ref{morphology} (top panel), the corresponding power-spectrum in
Fig. \ref{spectra} (top panel). 

With these turbulent initial conditions we perform two
simulations, one with and one without the inclusion of ionization. 
To account for the UV-radiation of a young massive star we use iVINE
(G08), a new parallel implementation
of ionizing 
radiation in the tree-SPH code VINE. Here we assume plane-parallel infall
of UV-radiation onto one transparent side 
of the simulated area, which enables us to
perform simulations at yet unmatched high resolution. From the
surface the radiation is propagated 
by a ray-shooting algorithm. The size of the rays is determined by the
smoothing-length of the SPH-particles, i.e. the width a particle
occupies. Along these rays the radiation is calculated.
This provides us with an ionization degree $\eta$ for each
SPH-particle, which is then used to assign 
a new temperature to each particle by linear interpolation.
\begin{equation}
T = T_\mathrm{hot} \cdot \eta + T_\mathrm{cold} \cdot (1-\eta),
\end{equation}
where $ T_\mathrm{cold} = 10$K is the initial temperature of the cold, unionized
gas and  $ T_\mathrm{hot} = 10^4$K is the average temperature of the ionized
gas \citep[see e.g.][]{1991pagd.book.....S}. The gas is assumed to be
atomic hydrogen.
Both gas components are close to thermal equilibrium since the
heating and cooling timescales are much shorter than the dynamical
timescales. We treat the gas with an 
isothermal equation of state ($\gamma=1$) as for the density range in
our simulations heating and cooling should balance each other to
approximate isothermality \citep[see
e.g.][]{1998ApJ...504..835S}. However, in reality the situation is
more complicated. Recent simulations by \cite{2007ApJS..169..239G}
indicate an equation of state of the thermal equilibrium gas which is
softer than isothermal ($\gamma=0.7-0.8$). 
For a detailed prescription of the iVINE-code
along with several analytical test cases see G08.
In the simulations presented here the radiation was calculated on more
than $(60)^2$ rays, with the additional inclusion of five levels of refinement,
leading to a spatial resolution of $2\times 10^{-3}$pc in the radiation.
The photon flux
per unit time and area is set to $F_{Ly}=5\times 10^9\textrm{
  photons }\textrm{cm}^{-2}\textrm{s}^{-1}$, 
allowing the radiation to penetrate the first $10\%$ of the cloud
immediately. This corresponds to setting up our simulation to be right
at the border of the Str{\"o}mgren-sphere
\citep{1939ApJ....89..526S}, which can be immediately ionized by an
O-star or association.

The radiation is impinging from the negative
x-direction. Hydrodynamics is calculated with periodic boundaries in
the y- and z-direction. The boundary is assumed to
be reflecting in the negative x-direction to represent conservation of
flux towards the star, whereas in the positive x-direction the gas is
allowed to stream away freely. Gravitational forces are calculated
without boundaries. This is valid as the free-fall time
of the whole simulated area is $ t_\mathrm{ff}\approx 3$Myr, which is
much larger than the simulation time. 
 To ensure a correct integration of all 
quantities we use the individual time-stepping-scheme of VINE with 
the same parameters as for the freely decaying turbulence (see
above). For the tree-based calculation of gravitational forces we use a
multi-pole acceptance  criterion (MAC, \citealt{2001NewA....6...79S})
with a tree accuracy parameter of $\theta=5\times10^{-4}$.
The correct treatment of the ionization and the resulting acceleration
of the particles is obtained by a modified CFL-condition as discussed
in G08.
The simulations are performed with $2\times10^6$  gas particles on a SGI Altix
3700 Bx2 supercomputer. The entire calculation took approximately 100
wall clock hours on 16 CPUs.

\section{Results}
\subsection{Morphology and Formation of Cores}
\label{morph}

At the beginning of the simulation the R-type front immediately reaches into the
first $10\%$ of the box, with the radiation penetrating further into
the low density parts of the cold gas.  After a hydrodynamical
crossing timescale of the hot gas
($t_\mathrm{hot}\approx30$kyr)\footnote{This timescale is calculated by
  taking the sound speed of the hot ionized gas
  $c_\mathrm{s,hot}=13.1\textrm{pc/Myr}$ and the average penetration
  length of the ionization of $0.4$pc into account.} the
ionized gas reacts to its increase in temperature and starts to exert
pressure on the cold gas. 
The cold gas is compressed and pushed away from the source,
leading to a systematic velocity in the x-direction.
At the same time the radiation has penetrated and ionized the ISM
along channels of low density gas. Now these low density regions
expand and start compressing the 
denser, unionized regions especially tangential to the direction of
radiation. Thus the preexisting density structures, which are seeded
by the turbulent initial conditions get enhanced as shown in
Fig. \ref{morphology}.

The combination of overall and tangential compression
leads to elongated structures that
keep sweeping up cold gas. After $\approx 250$kyr
(Fig. \ref{morphology}, middle panel) the dominant
structures are already excavated by the combination of radiation and
the pressure of the hot gas. From now on the evolution is mainly
dominated by the hydrodynamic interactions between the hot and cold
phase of the gas. 

After $\approx 500$kyr (Fig. \ref{morphology}, bottom panel) the
morphology is remarkably reminiscent of the observed structures. 
The pillars in our simulations are indeed very complex structures with
a cork-screw type, torqued morphology and show rotational motion
around their main axis, as it is observed
\citep{2006A&A...454..201G}. Up to now it has been suggested that
these complex morphologies arise due to magnetic fields, which are not
included in our simulations. It is very likely that the pillars in
M16 are a snapshot of the formation scenario proposed here. 
At this stage the densest region (indicated by the center of the white box in
Fig. \ref{morphology}, bottom panel) undergoes gravitational collapse,
the simulation is slowed down considerably and we terminate it. Future
simulations with the inclusion of e.g. sink 
particles to avoid the detailed calculation of the further gravitational
collapse leading to low mass stars will allow us to trace the subsequent evolution of the whole region.
We call the most prominent feature in the white box in
Fig. \ref{morphology} (bottom panel) 'pillar I' and the second largest 
'pillar II', the collapsing compact core is at the tip of pillar II. Their
respective masses are
$M_\mathrm{pillarI}=12.3\mathrm{M}_\odot$,
$M_\mathrm{pillarII}=8.1\mathrm{M}_\odot$ and
$M_\mathrm{core}=0.7\mathrm{M}_\odot$.
The compact core is defined as all material with a number density
above $n_\mathrm{crit}=10^7\textrm{cm}^{-3}$ in a region of
$R_\mathrm{crit}=0.02$pc around the peak density (see G08).
Observations show that star formation is taking place close to the tips of the
evolving structures \citep{2007arXiv0711.1515S}. The same is true for
our simulations. In the process of sweeping up the dense
material lags behind, gaining less momentum and thus leading to very
high density enhancements near the radiation front. In contrast, the
simulation without UV-radiation does not show any signs of
gravitational collapse.

Overall the scenario is very similar to the 'collect and collapse'
model, as the denser regions would not collapse on their own on the 
timescale simulated and the sweeping up of material plays a vital role. 
We call it 'collect and collapse with turbulent seeds'.

\subsection{Turbulent Evolution}
\label{specevol}

\begin{figure}
  \centering
\plotone{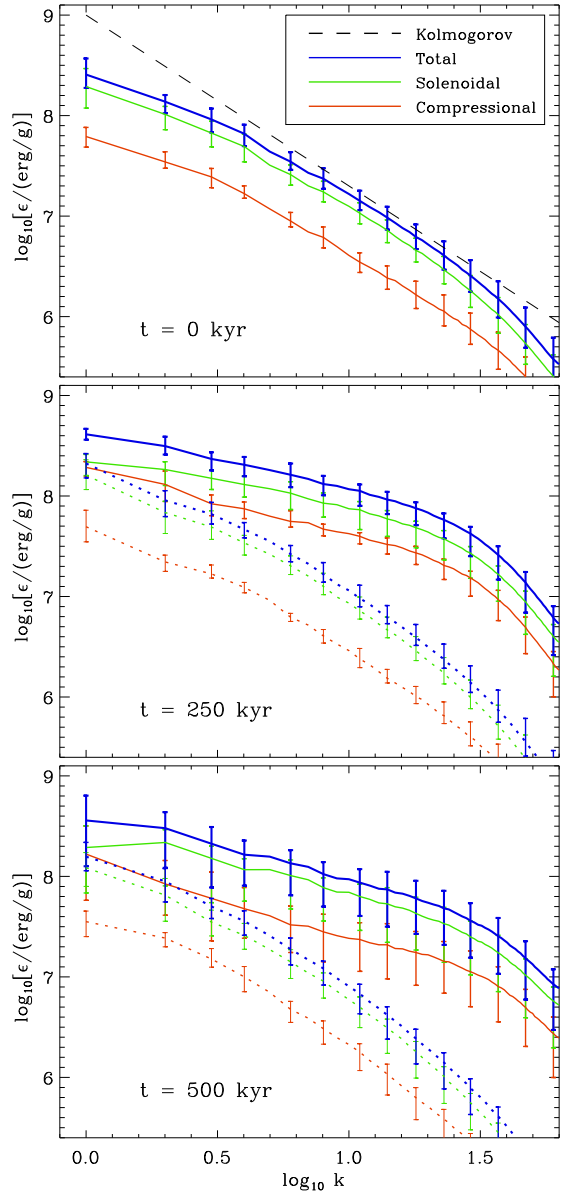}
 \caption{Evolution of the density weighted power
  spectra, plotted is the logarithm of the wavenumber versus the
  logarithm of the specific kinetic energy. Solid lines denote the run
  with ionization, dotted lines the control run without
  ionization. Plotted is the mean value of the spectra in the four
  different backward cubes, the error bars show the minimum 
  and maximum values. Blue: total specific kinetic energy, green:
  solenoidal modes, red: compressional modes, dashed: Kolmogorov
  power-law. }\label{spectra}
\end{figure}

For the discussion of the evolution of the density weighted spectra
from the run with ionization we perform a control run
with the same initial conditions and accuracy parameters
as the main simulation but without the inclusion of UV-radiation. In the
comparison run no sign of star formation can be
detected. This is reasonable, since the
cloud is not set up to become gravitationally unstable as the total time
of the simulation is much less than $t_\mathrm{ff}$.
The comparison is shown at $t=0$kyr, $t=250$kyr and $t=500$kyr
(Fig. \ref{spectra}). We analyze the turbulent spectrum in four cubic
domains in the backward domain of the simulation, spanning $2$pc in
each direction (one is indicated in Fig.\ref{morphology}, top panel, the
other three are shifted in the negative y- and z-direction,
respectively). Thus, it is guaranteed that there is always  
enough cold gas in the volume to achieve sensible results. To avoid a
bias by either the ionized gas or the forming high density regions we
take into account only gas 
with a number density $10^2 \textrm{cm}^{-3} < n < 10^4
\textrm{cm}^{-3}$. The particles are binned on a $(128)^3$ cubic
grid by using a kernel-weighted 
binning routine \citep[as e.g. in ][]{2008arXiv0810.4599K}. Based
on this grid we calculate the density weighted 
spectra by substituting
$\mathbf{v}$ with
$(\rho/\bar{\rho})^{1/2}(\mathbf{v}-\mathbf{v}_\mathrm{RMS})$ before
the Fourier transformation. $\mathbf{v}_\mathrm{RMS}$ is the average
velocity in each of the three components. The specific turbulent
kinetic energy in the Fourier space is then given as 
\begin{equation}
\epsilon_\mathrm{kin,turb}=\frac{1}{2}(\mathbf{v'}\cdot\mathbf{\bar{v'}}),
\end{equation}
where $\mathbf{v'}$ and $\mathbf{\bar{v'}}$ are the Fourier transform
of the substituted velocity and its complex conjugate, respectively.
By mapping the $\epsilon_\mathrm{kin,turb}$ cube to wave numbers
$\mathbf{k}$, the specific energy in the compressional, curl-free modes can
be calculated as  
\begin{equation}
\epsilon_\mathrm{com}=\epsilon_\mathrm{turb,kin}\frac{(\mathbf{v'}\cdot\mathbf{k})(\mathbf{\bar{v'}}\cdot\mathbf{k})}{(\mathbf{k}\cdot\mathbf{k})(\mathbf{v'}\cdot\mathbf{\bar{v'}})}.
\end{equation}
The specific energy in the solenoidal or incompressible, divergence-free modes
is then given by
$\epsilon_\mathrm{sol}=\epsilon_\mathrm{kin,turb}-\epsilon_\mathrm{com}$.
We construct the spectra by collecting the energy in the
different wavenumber intervals \citep[see][ for details]{2008arXiv0810.4599K}.
The total spectra as well as the solenoidal and compressional parts
are shown in Fig. \ref{spectra}.

The initial spectrum at $t=0$kyr (Fig.\ref{spectra}, top panel) resembles
quite well a power-law, even though the large-scale (low k) modes are
lower, as
the initial conditions are not produced by driven but by
freely decaying turbulence. The slope is similar to a
Kolmogorov-law. Approximately 25\% of the total turbulent energy is
contained in the compressional modes.

At $t=250$kyr (Fig.\ref{spectra}, middle panel) there is already a
distinct difference between the two spectra. The control run keeps the
power-law shape and  dissipates energy. In the ionization case 
the power is strongly increased 
on all scales. The increase is pronounced for $k>10$,
corresponding to scales $<0.2$pc. An interesting feature is the
rise in the compressional modes, where now 39\% of the total turbulent
energy is contained, whereas in the comparison run this ratio stays at
25\%. This clearly shows that the energy of the radiation is
transferred into compression of the cold gas via the hot gas. The
increase is in the turbulent energy itself and not correlated to the
overall bulk motion in the x-direction, since the mean velocity is
subtracted separately in each direction before the Fourier transform.

After $t=500$kyr (Fig.\ref{spectra}, bottom panel) these differences are
even more pronounced. 
The kinetic energy in the cold gas is now  a factor of four 
higher than in the run without ionization.
Approximately 33\% of the total turbulent energy is now contained in the
compressional modes. This suggest that after an initial phase of high
compression the system starts to relax. 

Including the mass in the respective region and density range the
total turbulent energy can be calculated. The initial turbulent energy
is $E_\mathrm{turb}= 2.1\times10^{45}\mathrm{erg}$, the final
turbulent energy (at $t=500$kyr) is $E_\mathrm{ion} =
4.3\times10^{45}\mathrm{erg}$ and 
$E_\mathrm{nion} = 1.1\times10^{45}\mathrm{erg}$ in the ionized and
unionized case, respectively. Thus, the input of turbulent energy per unit
volume and unit time averaged over the simulation time when comparing
the run with ionization to the case of freely decaying turbulence is
$\dot{e}_\mathrm{turb}=2.1\times10^{-25}\mathrm{erg}\,\mathrm{s}^{-1}\mathrm{cm}^{-3}$. By
using the simplified assumption that the UV-radiation is absorbed
isotropically in the entire simulation volume the amount of energy
contained in the ionizing radiation for the chosen flux
$F_\mathrm{Ly}$ is
$\dot{e}_\mathrm{Ly}=3.5\times10^{-20}\mathrm{erg}\,\mathrm{s}^{-1}\mathrm{cm}^{-3}$. Compared
to the estimates of \cite{2002ApJ...566..302M} and
\cite{2004RvMP...76..125M} our radiative energy is several orders of
magnitude higher, since we look at the direct surrounding of an O-star
instead of averaging over an entire galaxy. Nevertheless, the
conversion efficiency of ionization into turbulent motion of the cold
gas is in our case
$\sigma =
\dot{e}_\mathrm{turb}/\dot{e}_\mathrm{Ly}\approx2\times10^{-5}$,
which is an order of magnitude higher than their estimate of
$\sigma\approx2\times 10^{-6}$ for the Milky Way. Our highly resolved
simulations show that ionizing radiation from an O-star or association
provides a much more efficient mechanism to drive and sustain
turbulence in the parental molecular cloud than was previously
estimated. However, this is still the energy input into the local
environment in contrast to the average input rate on galactic scales
derived by \cite{2004RvMP...76..125M}. On the larger scales it does
not appear to contribute as significantly as e.g. supernova explosions.

\section{Discussion}
We have shown in this letter that the observed pillar-like structures
around O-stars as well as the gravitational collapse at the tip of the
pillars can result from the impact of the ionizing radiation of
massive O-stars on a turbulent molecular cloud. In addition, the
turbulent energy in the cold gas is increased by a factor of four,
especially in the compressional modes.  
Both effects are due to the same mechanism: the ionization can
heat the gas along channels of low density, thereby compressing gas at
higher density into filaments. Close to the source of ionization
this leads to the excavation of pillar-like structures with triggered
gravitational collapse at their tips. Further away from the source
front, the structures have not yet fully developed, nevertheless the
effect of compression is clearly visible in the turbulent
energy spectra.  

Even though we find striking similarities between our simulations and
observations, one has to bear in mind that this is a simplified
approach which does not involve full radiative transfer. Ionized gas which gets
shaded is assumed to cool immediately without affecting the adjacent
structures. In addition, the shaded gas does not get ionized and
heated by the recombination radiation of the ionized gas surrounding
it. This might influence the 
precise shape of the structure behind the tip. 
Moreover, the thin surface layers  around each pillar
where cold and hot gas are mixing cannot be resolved, although they might be
crucial for the precise understanding of the temperature and the
chemical composition of these structures.
Nevertheless, our simulations indicate that these detailed effects are
of minor importance to explain the global picture, i.e. the overall
structure and mass assembly of the pillars observed.
Stellar winds might have an additional impact. Although
O-stars have very powerful winds which can reach velocities of up to
$1000$km/s, our models suggest that ionizing radiation alone can
reproduce most observed features.

The straightforward combination of hydrodynamics and ionizing
radiation together with a standard turbulent model and
typical parameters for molecular clouds leads to morphologies
consistent with 
observed objects like pillars and collapsing cores. The
similarities suggest that ionizing radiation plays a major role not
only in shaping the parental cloud, but also in triggering
secondary star formation. Furthermore, the overall turbulent kinetic
energy in the cold gas is increased significantly.

\begin{acknowledgements}
We would like the thank the referee for the valuable comments which
helped to improve the manuscript.

This research was founded by the DFG cluster of excellence 'Origin
and Structure of the Universe'. 

All simulations were performed on a
SGI Altix 3700 Bx2 supercomputer that was partly funded by the DFG
cluster of excellence 'Origin and Structure of the Universe'.
\end{acknowledgements}

\bibliographystyle{apj}

\clearpage
\end{document}